\documentclass[12pt]{article}
\usepackage{a4wide}
\usepackage{latexsym}
\usepackage{amsmath}
\usepackage{amsfonts}
\usepackage{amscd}
\usepackage{cite}
\usepackage{placeins}

\usepackage{pslatex}
\usepackage{graphicx}
\usepackage[latin1,utf8]{inputenc}
\usepackage[T1]{fontenc}

\allowdisplaybreaks

\newcommand{\bq}{\begin{eqnarray}}
\newcommand{\eq}{\end{eqnarray}}
\newcommand{\eps}{\varepsilon}

\begin{document}

\thispagestyle{empty}

\begin{flushright}
  MITP/19-023
\end{flushright}

\vspace{1.5cm}

\begin{center}
  {\Large\bf Two-loop master integrals for the mixed QCD-electroweak corrections for $H \rightarrow b\bar{b}$ through a $H t \bar{t}$-coupling \\
  }
  \vspace{1cm}
  {\large Ekta Chaubey and Stefan Weinzierl \\
  \vspace{1cm}
      {\small \em PRISMA Cluster of Excellence, Institut f{\"u}r Physik, }\\
      {\small \em Johannes Gutenberg-Universit{\"a}t Mainz,}\\
      {\small \em D - 55099 Mainz, Germany}\\
  } 
\end{center}

\vspace{2cm}

\begin{abstract}\noindent
  {
We present the two-loop master integrals relevant to the ${\mathcal O}(\alpha \alpha_s)$-corrections 
to the decay $H \rightarrow b \bar{b}$ through a $H t \bar{t}$-coupling.
We keep the full dependence on the heavy particle masses, but neglect the $b$-quark mass.
The occurring square roots can be rationalised and the result is expressed in terms of multiple polylogarithms.
   }
\end{abstract}

\vspace*{\fill}

\newpage

\section{Introduction}
\label{sect:intro}

The precise determination of the properties of the Higgs boson is now a central pillar of the experimental programs at the LHC.
As the Higgs boson decays predominately to $b \bar{b}$, the partial width for this decay is of central importance \cite{Lepage:2014fla}.
This raises immediately the question, how precise can we predict this partial width within the Standard Model from theory?
This requires the computation of quantum corrections.

The state-of-the-art for the partial decay width for $H \rightarrow b \bar{b}$ is as follows:
In massless QCD the corrections are known to ${\mathcal O}(\alpha_s^4)$ \cite{Baikov:2005rw}.
Keeping the mass dependence, QCD corrections are known to ${\mathcal O}(\alpha_s^3)$ \cite{Chetyrkin:1995pd,Chetyrkin:1997mb,Chetyrkin:1997vj,Bernreuther:2018ynm}.
The QCD calculation of order ${\mathcal O}(\alpha_s^2)$
for the decay $H \rightarrow b \bar{b}$ through a $H t \bar{t}$-coupling
has become available quite recently \cite{Primo:2018zby}.
Two-loop QED corrections and mixed QED-QCD corrections have been considered in \cite{Kataev:1997cq}.
For the mixed electroweak-QCD corrections to the decay $H \rightarrow b \bar{b}$ of order ${\mathcal O}(\alpha \alpha_s)$
the leading term in an expansion in $m_H^2/m_t^2$ has been obtained in \cite{Kwiatkowski:1994cu,Kniehl:1994ju}.
This has been improved by including systematically more terms in \cite{Mihaila:2015lwa}.
As far as the two-loop electroweak corrections of order ${\mathcal O}(\alpha^2)$ are concerned only the leading
term in an expansion in $m_H^2/m_t^2$ is known \cite{Butenschoen:2006ns,Butenschoen:2007hz}.
In recent years, there has been a substantial progress in our abilities to compute Feynman integrals, 
and Feynman integrals which previously could only be calculated approximatively come into reach.
In this article we present 
the two-loop master integrals relevant to the ${\mathcal O}(\alpha \alpha_s)$-corrections to the decay $H \rightarrow b \bar{b}$ through a $H t \bar{t}$-coupling.
We keep the full dependence on the heavy particle masses ($m_t$, $m_H$ and $m_W$), but neglect the $b$-quark mass.
In Higgs and top physics there are two-loop Feynman integrals related to elliptic curves \cite{Bonciani:2016qxi,vonManteuffel:2017hms,Mistlberger:2018etf,Adams:2018bsn,Adams:2018kez}.
One might fear that we are in a similar situation here.
To some relief this is not the case.
All master integrals can be expressed entirely in terms of multiple polylogarithm.
The difficulty of these integrals is entirely due to the required simultaneous rationalisation of two square roots.

Large parts of the calculation are based on by now standard techniques:
We first derive a system of differential equations for the master integrals \cite{Kotikov:1990kg,Kotikov:1991pm,Remiddi:1997ny,Gehrmann:1999as,Argeri:2007up,MullerStach:2012mp,Henn:2013pwa,Henn:2014qga,Ablinger:2015tua,Adams:2017tga,Bosma:2017hrk}, 
using integration-by-parts identities \cite{Tkachov:1981wb,Chetyrkin:1981qh}
and the Laporta algorithm \cite{Laporta:2001dd}.
We then bring the system of differential equations into an $\eps$-form \cite{Henn:2013pwa}.
This will introduce two square roots.
It turns out that the square roots can be rationalised simultaneously with the methods of \cite{Besier:2018jen}.
This is the essential new ingredient of our calculation.
The resulting system of differential equations involves only $\mathrm{dlog}$-forms and can be solved in terms of multiple polylogarithms.
We are able to express all master integrals in terms of multiple polylogarithms with an alphabet consisting of $13$ letters.

This paper is organised as follows:
In the next section we introduce our notation.
In particular we define the kinematic variables which will rationalise the square roots.
In section~\ref{sect:masters} we define the master integrals, for which the system of differential equations is in $\eps$-form.
The system of differential equations is presented in section~\ref{sect:differential_equation}.
The analytic results for the master integrals are given in section~\ref{sect:analytical_results}.
In addition, section~\ref{sect:numerical_results} gives numerical results for the most interesting case $p^2=m_H^2$.
Finally, our conclusions are contained in section~\ref{sect:conclusions}.
The appendix shows for all master integrals the corresponding Feynman diagrams and describes the content of 
the supplementary electronic file attached to the arxiv version of this article.


\section{Notation}
\label{sect:notation}

\begin{figure}
\begin{center}
\includegraphics[scale=1.0]{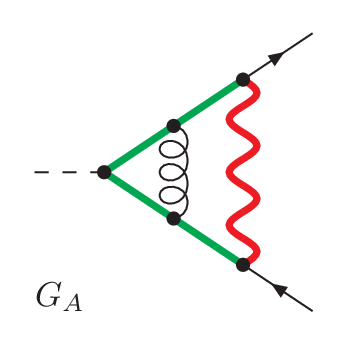}
\includegraphics[scale=1.0]{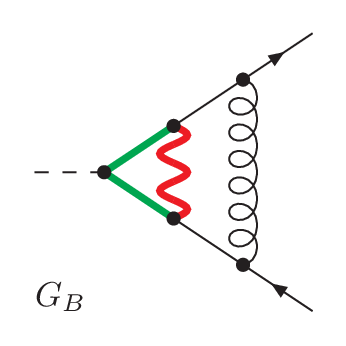}
\includegraphics[scale=1.0]{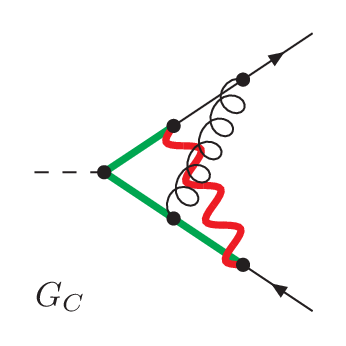}
\includegraphics[scale=1.0]{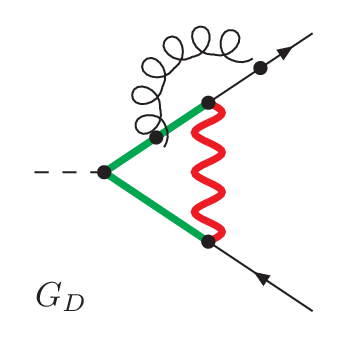}
\includegraphics[scale=1.0]{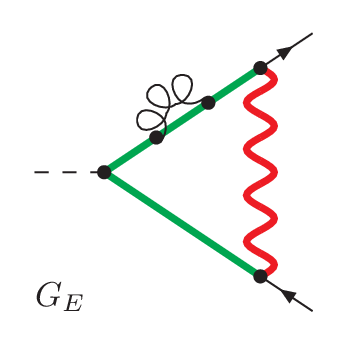}
\end{center}
\caption{
Examples of Feynman diagrams contributing to the mixed ${\mathcal O}(\alpha \alpha_s)$-corrections to the decay $H \rightarrow b \bar{b}$ through a $H t \bar{t}$-coupling.
The Higgs boson is denoted by a dashed line, a top quark by a green line, a bottom quark with a black line
and a gluon by a curly line.
Particles with mass $m_W$ are drawn with a wavy line.
}
\label{fig_Feynman_diagrams}
\end{figure}

We are interested in the mixed ${\mathcal O}(\alpha \alpha_s)$-corrections to the decay $H \rightarrow b \bar{b}$ 
through a $H t \bar{t}$-coupling.
Examples of Feynman diagrams are shown in fig.~\ref{fig_Feynman_diagrams}.
Not shown are diagrams whose master integrals are related to the master integrals of the diagrams 
of fig.~\ref{fig_Feynman_diagrams} by symmetry.
We will neglect the $b$-quark mass. 
However, we will treat the dependence on the top quark mass $m_t$, the $W$-boson mass $m_W$ and the momentum $p$ of the 
Higgs boson exactly.
For an on-shell Higgs-boson we have $p^2=m_H^2$.
For the two-loop contributions to the Higgs decay we have two independent external momenta $p_1$ and $p_2$, which label
the momenta of $b$-quark and $\bar{b}$-quark, respectively.
With two independent loop momenta we thus have seven linearly independent scalar products.
For each of the four Feynman diagrams $G_A$-$G_D$ we introduce an auxiliary topology with seven propagators.
\begin{figure}
\begin{center}
\includegraphics[scale=1.0]{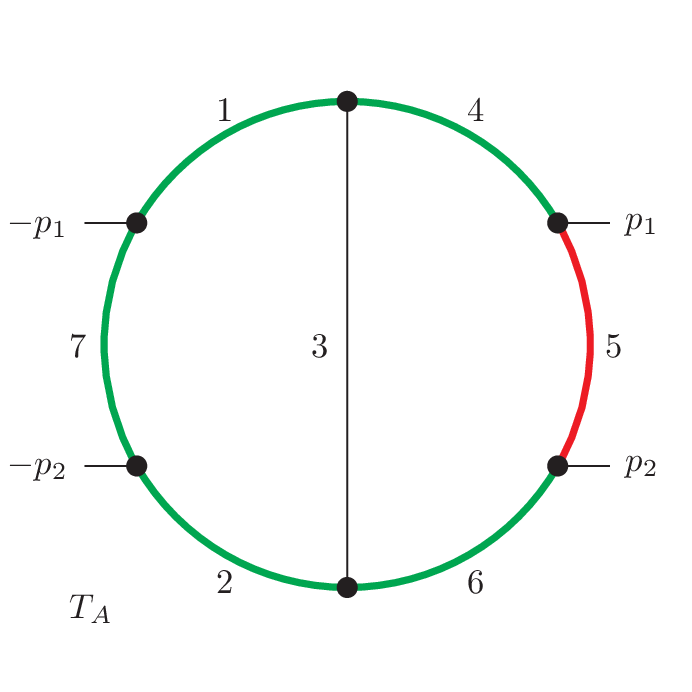}
\hspace*{10mm}
\includegraphics[scale=1.0]{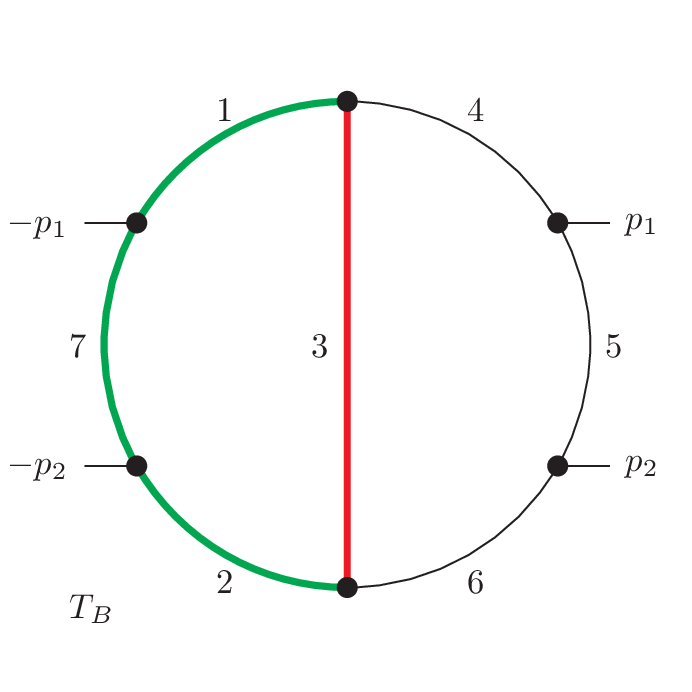}
\\
\includegraphics[scale=1.0]{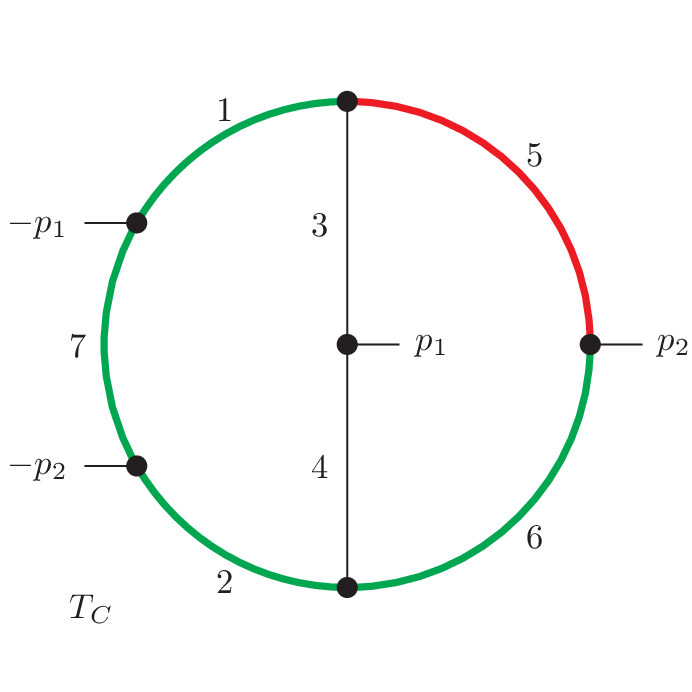}
\hspace*{10mm}
\includegraphics[scale=1.0]{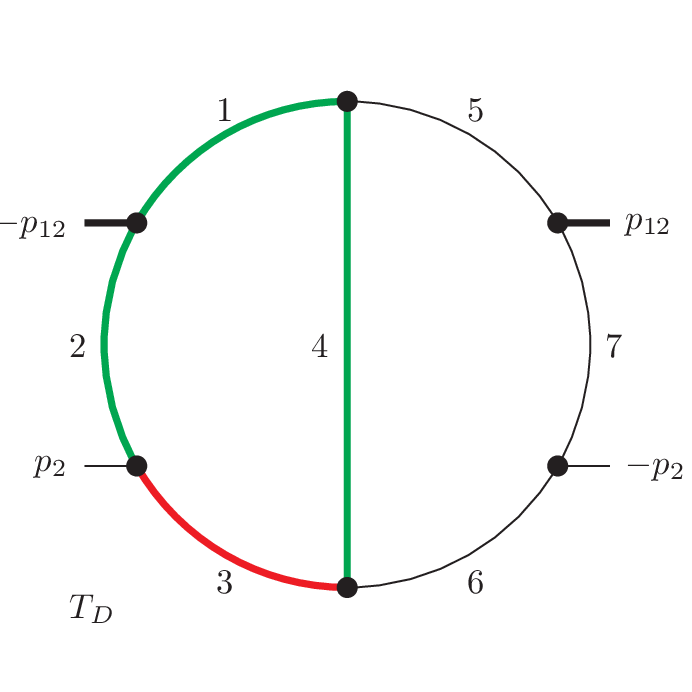}
\end{center}
\caption{
Auxiliary diagrams for the two planar vertex corrections, the non-planar vertex correction and topology $D$.
The internal masses of the propagators are encoded by the colour of the propagators: 
massless (black), $m_t$ (green), $m_W$ (red).
}
\label{fig_auxiliary_non_planar}
\end{figure}
These are shown in fig.~\ref{fig_auxiliary_non_planar}.
The master integrals related to diagram $G_E$ are a subset of the master integrals related to diagram $G_A$
(and likewise a subset of the master integrals related to diagram $G_D$).
We consider the integrals
\bq
\label{def_integral}
 I_{\nu_1 \nu_2 \nu_3 \nu_4 \nu_5 \nu_6 \nu_7}^X
 & = &
 e^{2 \gamma_E \eps}
 \left(\mu^2\right)^{\nu-D}
 \int \frac{d^Dk_1}{i \pi^{\frac{D}{2}}} \frac{d^Dk_2}{i \pi^{\frac{D}{2}}}
 \prod\limits_{j=1}^7 \frac{1}{ \left(P_j^X\right)^{\nu_j} },
 \;\;\;\;\;\;
 X \in \{A,B,C,D\},
\eq
where $D=4-2\eps$ denotes the number of space-time dimensions,
$\gamma_E$ denotes the Euler-Mascheroni constant, 
$\mu$ is an arbitrary scale introduced to render the Feynman integral dimensionless,
and the quantity $\nu$ is defined by
\bq
 \nu & = &
 \sum\limits_{j=1}^7 \nu_j.
\eq
The inverse propagators $P_j^X$ are defined as follows:
\begin{description}
\item{Topology $A$:}
\begin{align}
 P_1^A & = -k_1^2 + m_t^2,
 &
 P_2^A & = -\left(k_1-p_1-p_2\right)^2 + m_t^2,
 &
 P_3^A & = -\left(k_1+k_2\right)^2,
 \nonumber \\
 P_4^A & = -k_2^2 + m_t^2,
 &
 P_5^A & = -\left(k_2+p_1\right)^2 + m_W^2,
 &
 P_6^A & = -\left(k_2+p_1+p_2\right)^2 + m_t^2,
 \nonumber \\
 P_7^A & = -\left(k_1-p_1\right)^2 + m_t^2.
 & &
\end{align}
\item{Topology $B$:}
\begin{align}
 P_1^B & = -k_1^2 + m_t^2,
 &
 P_2^B & = -\left(k_1-p_1-p_2\right)^2 + m_t^2,
 &
 P_3^B & = -\left(k_1+k_2\right)^2 + m_W^2,
 \nonumber \\
 P_4^B & = -k_2^2,
 &
 P_5^B & = -\left(k_2+p_1\right)^2,
 &
 P_6^B & = -\left(k_2+p_1+p_2\right)^2,
 \nonumber \\
 P_7^B & = -\left(k_1-p_1\right)^2 + m_t^2.
 & &
\end{align}
\item{Topology $C$:}
\begin{align}
 P_1^C & = -k_1^2 + m_t^2,
 &
 P_2^C & = -\left(k_1-p_1-p_2\right)^2 + m_t^2,
 &
 P_3^C & = -\left(k_1+k_2\right)^2,
 \nonumber \\
 P_4^C & = -\left(k_1+k_2-p_1\right)^2,
 &
 P_5^C & = -k_2^2 + m_W^2,
 &
 P_6^C & = -\left(k_2+p_2\right)^2 + m_t^2,
 \nonumber \\
 P_7^C & = -\left(k_1-p_1\right)^2 + m_t^2.
 & &
\end{align}
\item{Topology $D$:}
\begin{align}
 P_1^D & = -k_1^2 + m_t^2,
 &
 P_2^D & = -\left(k_1-p_1-p_2\right)^2 + m_t^2,
 &
 P_3^D & = -\left(k_1-p_1\right)^2 + m_W^2,
 \nonumber \\
 P_4^D & = -\left(k_1+k_2\right)^2 + m_t^2,
 &
 P_5^D & = -k_2^2,
 &
 P_6^D & = -\left(k_2+p_1\right)^2,
 \nonumber \\
 P_7^D & = -\left(k_2+p_1+p_2\right)^2.
 & &
\end{align}
\end{description}
For all topologies our conventions are such that we are interested in the integrals with $\nu_7 \le 0$.

The Feynman parameter representations for the four topologies are given by
\bq
 I_{\nu_1 \nu_2 \nu_3 \nu_4 \nu_5 \nu_6 \nu_7}^X
 & = &
 e^{2 \gamma_E \eps}
 \frac{\Gamma(\nu-D)}{\prod\limits_{j=1}^{7}\Gamma(\nu_j)}
 \;
 \int\limits_{x_i \ge 0}
 d^7x
 \;
 \delta\left(1-\sum\limits_{j=1}^7 x_j\right)
 \; 
 \left( \prod\limits_{j=1}^{7} x_j^{\nu_j-1} \right)\,\frac{{\mathcal U}_X^{\nu-\frac{3}{2}D}}
 {{\mathcal F}_X^{\nu-D}},
\eq
The graph polynomials are given by
\bq
\label{def_U_and_F}
 {\mathcal U}_A 
 & = & 
 \left( x_1+x_2+x_7 \right) \left( x_4+x_5+x_6 \right) + x_3 \left( x_1+x_2 + x_4+x_5+x_6+x_7\right),
 \nonumber \\
 {\mathcal F}_A 
 & = & 
 \left[ 
        x_1 x_2 \left(x_3+x_4+x_5+x_6\right)
        + x_4 x_6 \left(x_1+x_2+x_3+x_7\right)
        + x_3 \left( x_1 x_6 + x_2 x_4 \right) 
 \right] \left(\frac{-p^2}{\mu^2}\right)
 \nonumber \\
 & &
 + {\mathcal U}_A \left[ \left( x_1+x_2+x_4+x_6+x_7\right) \frac{m_t^2}{\mu^2} + x_5 \frac{m_W^2}{\mu^2} \right],
 \nonumber \\
 {\mathcal U}_B
 & = & 
 \left( x_1+x_2+x_7 \right) \left( x_4+x_5+x_6 \right) + x_3 \left( x_1+x_2 + x_4+x_5+x_6+x_7\right),
 \nonumber \\
 {\mathcal F}_B 
 & = & 
 \left[ 
        x_1 x_2 \left(x_3+x_4+x_5+x_6\right)
        + x_4 x_6 \left(x_1+x_2+x_3+x_7\right)
        + x_3 \left( x_1 x_6 + x_2 x_4 \right) 
 \right] \left(\frac{-p^2}{\mu^2}\right)
 \nonumber \\
& &
 + {\mathcal U}_B \left[ \left( x_1+x_2+x_7\right) \frac{m_t^2}{\mu^2} + x_3 \frac{m_W^2}{\mu^2} \right],
 \nonumber \\
 {\mathcal U}_C 
 & = & 
 \left( x_1+x_2+x_7 \right) \left( x_5+x_6 \right) + \left( x_3+x_4 \right) \left( x_1+x_2 + x_5+x_6 +x_7\right),
 \nonumber \\
 {\mathcal F}_C 
 & = & 
 \left[ 
        x_1 x_2 \left( x_3 + x_4 + x_5 + x_6 \right) + x_1 x_4 x_6 + x_2 x_3 x_5 - x_3 x_6 x_7
 \right] \left(\frac{-p^2}{\mu^2}\right)
 \nonumber \\
& &
 + {\mathcal U}_C \left[ \left( x_1+x_2+x_6+x_7\right) \frac{m_t^2}{\mu^2} + x_5 \frac{m_W^2}{\mu^2} \right],
 \nonumber \\
 {\mathcal U}_D 
 & = & 
 \left( x_1+x_2+x_3 \right) \left( x_5+x_6+x_7 \right) + x_4 \left( x_1+x_2+x_3 + x_5+x_6+x_7\right),
 \nonumber \\
 {\mathcal F}_D 
 & = & 
 \left[ 
        x_1 x_2 \left(x_4+x_5+x_6+x_7\right)
        + x_5 x_7 \left(x_1+x_2+x_3+x_4\right)
        + x_4 \left( x_1 x_7 + x_2 x_5 \right) 
 \right] \left(\frac{-p^2}{\mu^2}\right)
 \nonumber \\
& &
 + {\mathcal U}_D \left[ \left( x_1+x_2+x_4\right) \frac{m_t^2}{\mu^2} + x_3 \frac{m_W^2}{\mu^2} \right].
\eq
Let us introduce an operator ${\bf i}^+$, which raises the power of the propagator $i$ by one, e.g.
\bq
 {\bf 1}^+ I_{\nu_1 \nu_2 \nu_3 \nu_4 \nu_5 \nu_6 \nu_7}^X
 & = &
 I_{(\nu_1+1) \nu_2 \nu_3 \nu_4 \nu_5 \nu_6 \nu_7}^X.
\eq
In addition we define two operators ${\bf D}^\pm$, which shift the dimension of space-time by two through
\bq
 {\bf D}^\pm I_{\nu_1 \nu_2 \nu_3 \nu_4 \nu_5 \nu_6 \nu_7}^X\left( D \right)
 & = &
 I_{\nu_1 \nu_2 \nu_3 \nu_4 \nu_5 \nu_6 \nu_7}^X\left( D\pm 2 \right).
\eq
The dimensional shift relations read \cite{Tarasov:1996br,Tarasov:1997kx}
\bq
\label{dim_shift_eq}
 {\bf D}^- I_{\nu_1 \nu_2 \nu_3 \nu_4 \nu_5 \nu_6 \nu_7}^X\left(D\right)
 & = &
 {\mathcal U}_X\left( \nu_1 {\bf 1}^+, \nu_2 {\bf 2}^+, \nu_3 {\bf 3}^+, \nu_4 {\bf 4}^+, \nu_5 {\bf 5}^+ , \nu_6 {\bf 6}^+, \nu_7 {\bf 7}^+ \right)
 I_{\nu_1 \nu_2 \nu_3 \nu_4 \nu_5 \nu_6 \nu_7}^X\left(D\right).
 \nonumber \\
\eq
In the following we will set
\bq
 \mu^2 & = & m_t^2.
\eq
After setting $\mu^2=m_t^2$ the master integrals depend kinematically on two dimensionless quantities.
A naive choice is
\bq
 v \; = \; \frac{p^2}{m_t^2},
 & &
 w \; = \; \frac{m_W^2}{m_t^2},
\eq
with $p=p_1+p_2$.
However, with this choice we will encounter square roots.
In particular, the square roots
\bq
 \sqrt{-v\left(4-v\right)}
 & \mbox{and} &
 \sqrt{\lambda\left(v,w,1\right)}
\eq
will occur.
The K\"allen function is defined by
\bq
 \lambda(x,y,z) & = & x^2 + y^2 + z^2 - 2 x y - 2 y z - 2 z x.
\eq
In order to rationalise the square roots we introduce dimensionless quantities $x$ and $y$ through
\bq
\label{variable_trafo}
 \frac{p^2}{m_t^2} \; = \; v \; = \; - \frac{\left(1-x\right)^2}{x},
 & &
 \frac{m_W^2}{m_t^2} \; = \; w \; = \; \frac{\left(1-y+2xy\right)\left(x-2y+xy\right)}{x\left(1-y^2\right)}.
\eq
The first transformation is standard and has occurred in many places before, the second one is easily obtained
with the methods of ref.~\cite{Besier:2018jen}.
The Feynman integrals are then functions of $x,y$ and the dimensional regularisation parameter $\eps$.
The inverse transformations are given by
\bq
 x \; = \; \frac{1}{2} \left( 2-v - \sqrt{-v \left(4-v\right)} \right),
 & & 
 y \; = \;
 \frac{\sqrt{ \lambda\left(v,w,1\right) } - \sqrt{-v\left(4-v\right)}}{1-w+2v},
\eq
such that $x=0$ corresponds to $v=\infty$ and
$y=0$ corresponds to $w=1$.
Let us also note that the point $(v,w)=(0,1)$ is blown up in $(x,y)$-space to the hypersurface
$x=1$.
This motivates our final change of coordinates and we introduce
\bq
 x' & = & 1 - x.
\eq


\section{Master integrals}
\label{sect:masters}

For the reduction to master integrals we use the programs 
{\tt Reduze} \cite{vonManteuffel:2012np},
{\tt Kira} \cite{Maierhoefer:2017hyi} or
{\tt Fire} \cite{Smirnov:2014hma} combined with
{\tt LiteRed} \cite{Lee:2012cn,Lee:2013mka}.
Each topology involves a certain number of master integrals.
\begin{table}
\begin{center}
\begin{tabular}{|c|c|}
\hline
 Topology & Number of master integrals \\
\hline
 $A$ & $18$ \\
 $B$ & $15$ \\
 $C$ & $31$ \\
 $D$ & $14$ \\
\hline
\end{tabular}
\end{center}
\caption{The number of master integrals for a given topology.
}
\label{table_number_masters}
\end{table}
This number is shown in table~\ref{table_number_masters} and corresponds to the number of master integrals if we just consider one topology in isolation.
Of course, the various topologies share some master integrals and the final number of master integrals which we have to compute is lower.
We have the following relations
\begin{alignat}{4}
 & I_{\mu00\nu000}^A & \;\;\; = \;\;\; & I_{\mu0000\nu0}^C & \;\;\; = \;\;\; & I_{\mu00\nu000}^D, & &
 \nonumber \\
 & I_{\mu000\nu00}^A & \;\;\; = \;\;\; & I_{\mu0\nu0000}^B & \;\;\; = \;\;\; & I_{\mu000\nu00}^C & \;\;\; = \;\;\; & I_{00\nu\mu000}^D,
 \nonumber \\
 & I_{\mu\nu0\rho000}^A & \;\;\; = \;\;\; & I_{\mu\nu000\rho0}^C & \;\;\; = \;\;\; & I_{\mu\nu0\rho000}^D, & &
 \nonumber \\
 & I_{\mu\nu00\rho00}^A & \;\;\; = \;\;\; & I_{\mu\nu\rho0000}^B & \;\;\; = \;\;\; & I_{\mu\nu00\rho00}^C, & &
 \nonumber \\
 & I_{0\mu\nu\rho000}^A & \;\;\; = \;\;\; & I_{\mu00\nu0\rho0}^C & \;\;\; = \;\;\; & I_{0\mu0\rho\nu00}^D, & &
 \nonumber \\
 & I_{0\mu\nu\rho000}^B & \;\;\; = \;\;\; & I_{0\mu\rho0\nu00}^C, & & & &
 \nonumber \\
 & I_{\mu0\nu0\rho00}^A & \;\;\; = \;\;\; & I_{\mu0\rho\nu000}^B & \;\;\; = \;\;\; & I_{\mu0\nu0\rho00}^C & \;\;\; = \;\;\; & I_{00\rho\mu\nu00}^D,
 \nonumber \\
 & I_{\mu00\nu\rho\sigma0}^A & \;\;\; = \;\;\; & I_{\nu\sigma\rho\mu000}^D, & & & &
 \nonumber \\
 & I_{\mu\nu\rho\sigma000}^B & \;\;\; = \;\;\; & I_{\mu\nu\sigma0\rho00}^C, & & & &
 \nonumber \\
 & I_{\mu\nu\rho00\sigma0}^C & \;\;\; = \;\;\; & I_{\mu\nu0\sigma0\rho0}^D, & & & &
 \nonumber \\
 & I_{\mu\nu\rho0\sigma00}^A & \;\;\; = \;\;\; & I_{\mu\nu\sigma0\rho00}^B & \;\;\; = \;\;\; & I_{\mu\nu0\rho\sigma00}^C, & &
 \nonumber \\
 & I_{0\mu\nu\rho\sigma00}^A & \;\;\; = \;\;\; & I_{\mu00\nu\sigma\rho0}^C & \;\;\; = \;\;\; & I_{0\rho\sigma\mu\nu00}^D, & &
 \nonumber \\
 & I_{\mu\nu\rho\sigma\kappa00}^A & \;\;\; = \;\;\; & I_{\mu\nu0\rho\kappa\sigma0}^C, & & & &
 \nonumber \\
 & I_{\mu\nu\rho\sigma\kappa00}^B & \;\;\; = \;\;\; & I_{\mu\nu\sigma\kappa\rho00}^C, & & & &
 \nonumber \\
 & I_{\mu0\nu\rho\sigma\kappa0}^C & \;\;\; = \;\;\; & I_{0\kappa\sigma\mu\nu\rho0}^D. & & & &
\end{alignat}
In total we have to consider 39 master integrals, which are grouped into 25 blocks such that one block corresponds to one sub-topology.
Some of the master integrals are taken as integrals in $D-2=2-2\eps$ space-time dimensions.
Of course, with the help of the dimensional shift relations in eq.~(\ref{dim_shift_eq})
they are easily expressed as (longer) linear combinations of master integrals in $D=4-2\eps$ space-time dimensions.
A system of master integrals is given by
\bq
 J_{1}
 & = &
 \eps^2 \; {\bf D}^- I_{1001000}^A,
 \nonumber \\
 J_{2}
 & = &
 \eps^2 \; {\bf D}^- I_{1000100}^A,
 \nonumber \\
 J_{3}
 & = &
 \eps^2 \frac{\left(1-x^2\right)}{2x} \; {\bf D}^- I_{1101000}^A,
 \nonumber \\
 J_{4}
 & = &
 \eps^2 \frac{\left(1-x^2\right)}{2x} \; {\bf D}^- I_{1100100}^A,
 \nonumber \\
 J_{5}
 & = &
 \frac{1}{2} \eps^2 v \; {\bf D}^- I_{1001010}^B,
 \nonumber \\
 J_{6}
 & = &
 \frac{1}{2} \eps^2 v \; {\bf D}^- I_{0011010}^B,
 \nonumber \\
 J_{7}
 & = &
 \eps^2 \frac{\left(1-x^2\right)}{2x} \; {\bf D}^- I_{0111000}^A,
 \nonumber \\
 J_{8}
 & = &
 \eps^2 \; {\bf D}^- I_{(-1)111000}^A,
 \nonumber \\
 J_{9}
 & = &
 \eps^2 \frac{\left(1-x\right)\left(1+x-2y+y^2+2xy+xy^2\right)}{x\left(1-y^2\right)} \; {\bf D}^- I_{0111000}^B,
 \nonumber \\
 J_{10}
 & = &
 \eps^2 \left[ {\bf D}^- I_{(-1)111000}^B - \left(1-w\right) {\bf D}^- I_{0111000}^B \right],
 \nonumber \\
 J_{11}
 & = &
 2 \eps^2 \; {\bf D}^- I_{0111(-1)00}^B
 - 2 \eps^2 \; {\bf D}^- I_{0110000}^B,
 \nonumber \\
 J_{12}
 & = &
 \eps^2 \left(1-w\right) \; {\bf D}^- I_{1010100}^A,
 \nonumber \\
 J_{13}
 & = &
 \eps^2 \frac{\left(1-x^2\right)^2}{4x^2} \; {\bf D}^- I_{1101010}^A,
 \nonumber \\
 J_{14}
 & = &
 \eps^2 \frac{\left(1-x\right)^3\left(1+x\right)}{4x^2} \; {\bf D}^- I_{1101010}^B,
 \nonumber \\
 J_{15}
 & = &
 \eps^3 \left(1-\eps\right) v \; I_{1001110}^A,
 \nonumber \\
 J_{16}
 & = &
 \eps^2 \frac{\left(1-x^2\right)}{x} 
   \left[ \left(1-w\right) \; {\bf D}^- I_{1111000}^B
          - {\bf D}^- I_{0111000}^B
 \right],
 \nonumber \\
 J_{17}
 & = &
 \eps^3 v \; I_{1120010}^C,
 \nonumber \\
 J_{18}
 & = &
 \eps^3 v \; I_{1110020}^C,
 \nonumber \\
 J_{19}
 & = &
 \eps^2 \frac{\left(1-x^2\right)}{x} \left[
 \left(1-2\eps\right) \; I_{2110010}^C + \eps \; I_{1110020}^C \right],
 \nonumber \\
 J_{20}
 & = &
 \eps^3 v \; I_{1120100}^A,
 \nonumber \\
 J_{21}
 & = &
 \eps^3 v \; I_{1110200}^A,
 \nonumber \\
 J_{22}
 & = &
 2 \eps^2
 \frac{\left(1-x\right)}{x\left(1+w\right)}
 \left[
 \left(1-2\eps\right) \left(1+x\right) \; I_{2110100}^A
 - \eps \left(x-w\right) \; I_{1120100}^A
 + \eps \left(1-x+2w\right) \; I_{1110200}^A
 \right. \nonumber \\
 & & \left.
 - \frac{1}{2} \left(1+x\right) \; {\bf D}^- I_{1100100}^A
 \right] ,
 \nonumber \\
 J_{23}
 & = &
 2 \eps^3 v \; I_{0211100}^A,
 \nonumber \\
 J_{24}
 & = &
 2 \eps^3 v \; I_{0121100}^A,
 \nonumber \\
 J_{25}
 & = &
 2 \eps^3 v \; I_{0210110}^C,
 \nonumber \\
 J_{26}
 & = &
 2 \eps^3 v \; I_{0120110}^C,
 \nonumber \\
 J_{27}
 & = &
 \eps^3 \frac{\left(1-x\right)^2}{x\left(1+x\right)} 
 \left[
   \left(1-2\eps\right) \left(1-x\right) \; I_{1101110}^A
 + 2 \left(1-\eps\right) \; I_{1001110}^A 
 \right],
 \nonumber \\
 J_{28}
 & = &
 2 \eps^3 \left(1-2\eps\right) v \; I_{1111010}^B,
 \nonumber \\
 J_{29}
 & = &
 2 \eps^3 v \left(1-w\right) \; I_{1111200}^A,
 \nonumber \\
 J_{30}
 & = &
 2 \eps^3 v
   \left( I_{1112100}^A + I_{1111200}^A \right),
 \nonumber \\
 J_{31}
 & = &
 4 \eps^3 \frac{\left(1-x\right)^2}{x\left(1+x^2\right)} 
  \left[ 
         \left(1-x^2\right) I_{2111100}^A 
         - \left(1+w\right) I_{1111200}^A 
         - 2 I_{1112100}^A 
         + 2 I_{1120100}^A 
         +   I_{1110200}^A 
 \right. \nonumber \\
 & & \left.
         -   I_{0211100}^A 
         - 2 I_{0121100}^A 
  \right],
 \nonumber \\
 J_{32}
 & = &
 4 \eps^4 v \; I_{1111100}^B,
 \nonumber \\
 J_{33}
 & = &
 2 \eps^3 v \left(1-w\right) \; I_{1110120}^C,
 \nonumber \\
 J_{34}
 & = &
 2 \eps^3 v w 
   \left( I_{1110210}^C + I_{1110120}^C \right),
 \nonumber \\
 J_{35}
 & = &
 2 \eps^3 v \left(1-w\right) \; I_{1120110}^C,
 \nonumber \\
 J_{36}
 & = &
 4 \eps^3 \frac{\left(1-x\right)^2}{x\left(1+x^2\right)} 
  \left[ 
         \left(1-x^2\right) I_{2110110}^C 
         - \left(1+w\right) I_{1110120}^C 
         - 2 w \; I_{1110210}^C 
         + 2 \left(1-w\right) I_{1120110}^C 
 \right. \nonumber \\
 & & \left.
         + 2 I_{1120010}^C 
         +   I_{1110020}^C 
         -   I_{0210110}^C 
         - 2 I_{0120110}^C 
  \right],
 \nonumber \\
 J_{37}
 & = &
 2 \eps^4 v \; I_{1011110}^C,
 \nonumber \\
 J_{38}
 & = &
 \eps^3 v w \; I_{1112010}^D,
 \nonumber \\
 J_{39}
 & = &
 4 \eps^4 \frac{\left(1-x\right)^2\left(1-x+x^2-x w\right)}{x^2} \; I_{1111110}^C.
\eq


\section{The system of differential equations}
\label{sect:differential_equation}

Let us set $\vec{J}=(J_1,...,J_{39})^T$ for the vector of master integrals.
In this basis the system of differential equations is in $\eps$-form \cite{Henn:2013pwa}.
We have
\bq
 d \vec{J}
 & = &
 \eps A \vec{J},
\eq
where the matrix $A$ is independent of $\eps$.
In order to construct the basis of master integrals leading to an $\eps$-form,
we could have followed the same procedure as in \cite{Adams:2018bsn,Adams:2018kez}.
In practice, the results of \cite{Adams:2018bsn,Adams:2018kez} allow us to guess the master integrals 
or at least indicate an ansatz for the master integrals. 
In each topology, our ansatz involves only a few parameters, which are fixed by requiring that
terms of order $\eps^0$ in the matrix $A$ vanish.

Let us describe the singularities of the system of differential equations.
The singularities are on hypersurfaces and each hypersurface is defined by a polynomial in $x$ and $y$.
There are sixteen polynomials, which are given by 
\bq
 p_1
 & = &
 x,
 \nonumber \\
 p_2
 & = &
 x - 1,
 \nonumber \\
 p_3
 & = &
 x + 1,
 \nonumber \\
 p_4
 & = &
 y,
 \nonumber \\
 p_5
 & = &
 y - 1,
 \nonumber \\
 p_6
 & = &
 y + 1,
 \nonumber \\
 p_7
 & = &
 x y + x - y + 1,
 \nonumber \\
 p_8
 & = &
 x y + x - 2 y,
 \nonumber \\
 p_9
 & = &
 2 x y - y + 1,
 \nonumber \\
 p_{10}
 & = &
 x y^2 + 2 x y - 2 y^2 + x + 2 y,
 \nonumber \\
 p_{11}
 & = &
 x y^2 + 2 x y + y^2 + x - 2 y + 1,
 \nonumber \\
 p_{12}
 & = &
 x y^2 + 2 x y - y^2 + x + 2 y - 1,
 \nonumber \\
 p_{13}
 & = &
 2 x y^2 + 2 x y - y^2 + 2 y - 1,
 \nonumber \\
 p_{14}
 & = &
 2 x y^2 + 2 x y - 3 y^2 + 2 y + 1,
 \nonumber \\
 p_{15}
 & = &
 3 x y^2 + 2 x y - 2 y^2 - x + 2 y,
 \nonumber \\
 p_{16}
 & = &
 3 x y^2 + 2 x y - 3 y^2 - x + 2 y + 1.
\eq
We note that the polynomials $p_k$ are maximally of degree $3$.
The highest degree in the variable $y$ is two, the highest degree in the variable $x$ is one.
The entries of the matrix $A$ are ${\mathbb Q}$-linear combinations of $\mathrm{dlog}$-forms 
of these polynomials:
\bq
 A_{ij} & = 
 \sum\limits_{k=1}^{16} \tilde{c}_{i j k} \; d\ln\left(p_k\left(x,y\right)\right),
 \;\;\;\;\;\;
 \tilde{c}_{i j k} \; \in \; {\mathbb Q}.
\eq
We find that the matrix $A$ contains only fifteen
${\mathbb Q}$-independent linear combinations of $\mathrm{dlog}$-forms.
A basis for these is given by
\bq
\label{diff_forms}
 \omega_1
 & = & 
 \frac{ds}{s}
 \; = \;
 2 d \ln p_2 - d\ln p_1,
 \nonumber \\
 \omega_2
 & = &
 \frac{ds}{s-4m_t^2}
 \; = \;
 2 d \ln p_3 - d\ln p_1,
 \nonumber \\
 \omega_3
 & = &
 \frac{ds}{\sqrt{-s\left(4m_t^2-s\right)}}
 \; = \;
 d \ln p_1,
 \nonumber \\
 \omega_4
 & = &
 \frac{dm_W^2}{m_W^2}
 \; = \;
 d \ln p_8 + d \ln p_9 - d \ln p_5 - d \ln p_6 - d \ln p_1,
 \nonumber \\
 \omega_5
 & = &
 \frac{dm_W^2}{m_W^2-m_t^2}
 \; = \;
 d \ln p_7 - d \ln p_5 - d \ln p_6 + d \ln p_4 + d \ln p_2 - d \ln p_1,
 \nonumber \\
 \omega_6
 & = &
 d \ln\left(s + m_W^2-m_t^2\right)
 \; = \;
 d \ln p_{16} - d \ln p_5 - d \ln p_6 + d \ln p_2 - d \ln p_1,
 \nonumber \\
 \omega_7
 & = &
 d \ln\left(s - m_W^2+m_t^2\right)
 \; = \;
 d \ln p_{12} - d \ln p_5 - d \ln p_6 + d \ln p_2 - d \ln p_1,
 \nonumber \\
 \omega_8
 & = &
 \frac{1}{2} d \ln\left(s m_W^2 +\left(m_t^2-m_W^2\right)^2\right)
 \nonumber \\
 & = &
 \frac{1}{2} d \ln p_{15} + \frac{1}{2} d \ln p_{14} - d \ln p_5 - d \ln p_6 + d \ln p_2 - d \ln p_1,
 \nonumber \\
 \omega_9
 & = &
 \frac{1}{2} d \ln\left(\left(2 m_t^2 - m_W^2 \right) s +\left(m_t^2-m_W^2\right)^2\right)
 \nonumber \\
 & = &
 \frac{1}{2} d \ln p_{13} + \frac{1}{2} d \ln p_{10} - d \ln p_5 - d \ln p_6 + d \ln p_2 - d \ln p_1,
 \nonumber \\
 \omega_{10}
 & = &
 \frac{1}{2} d \ln\left(\lambda\left(s,m_W^2,m_t^2\right)\right)
 \nonumber \\
 & = &
 d \ln p_{11} - d \ln p_5 - d \ln p_6 + d \ln p_2 - d \ln p_1,
 \nonumber \\
 \omega_{11}
 & = &
 \frac{1}{2} d \ln p_{15} - \frac{1}{2} d \ln p_{14},
 \nonumber \\
 \omega_{12}
 & = &
 \frac{1}{2} d \ln p_{13} - \frac{1}{2} d \ln p_{10},
 \nonumber \\
 \omega_{13}
 & = &
 \frac{1}{2} d \ln p_{9} - \frac{1}{2} d \ln p_{8},
 \nonumber \\
 \omega_{14}
 & = &
 \frac{1}{2} d \ln p_{6} - \frac{1}{2} d \ln p_{5},
 \nonumber \\
 \omega_{15}
 & = &
 d \ln p_4 - \frac{1}{2} d \ln p_{6} - \frac{1}{2} d \ln p_{5}.
\eq
The entries of $A$ are therefore of the form
\bq
 A_{ij} & = 
 \sum\limits_{k=1}^{15} c_{i j k} \; \omega_k,
 \;\;\;\;\;\;
 c_{i j k} \; \in \; {\mathbb Q}.
\eq
By a rescaling of the master integrals with constant factors we may actually achieve
\bq
 c_{i j k} \; \in \; {\mathbb Z}.
\eq
For our choice of basis of master integrals $\vec{J}$ we have $c_{i j k} \in {\mathbb Z}$.
Equivalently, we may express the matrix $A$ as
\bq
 A & = &
 \sum\limits_{k=1}^{15} C_k \; \omega_k,
\eq
where the entries of the $39 \times 39$-matrices $C_k$ are integer numbers.
The matrix $A$ is given in the supplementary electronic file attached to the arxiv version of this article.

On specific hypersurfaces the differential forms simplify considerably.
On the hypersurface $y=0$ (i.e. for the case $m_W=m_t$) the differential forms reduce to a linear combination of
\bq
 \frac{dx}{x},
 \;\;\;
 \frac{dx}{x-1},
 \;\;\;
 \frac{dx}{x+1}.
\eq
On the hypersurface $x=0$ (i.e. for the case $p^2 \rightarrow \infty$) the differential forms reduce to a linear combination of
\bq
 \frac{dy}{y},
 \;\;\;
 \frac{dy}{y-1},
 \;\;\;
 \frac{dy}{y+1},
 \;\;\;
 \frac{dy}{y+\frac{1}{3}}.
\eq
On the hypersurface $x=1$ (i.e. for the case $p^2=0$ and $m_W^2=m_t^2$) the differential forms reduce to a linear combination of
\bq
\label{integration_kernels_xeq1}
 \frac{dy}{y},
 \;\;\;
 \frac{dy}{y-1},
 \;\;\;
 \frac{dy}{y+1},
 \;\;\;
 \frac{2 y dy}{y^2+1},
 \;\;\;
 \frac{2 \left(y-2\right) dy}{y^2-4y-1},
 \;\;\;
 \frac{2 \left(y+2\right) dy}{y^2+4y-1}.
\eq
The derivative of the master integrals is given by the product of the matrix $A$ with the vector $\vec{J}$:
\bq
 d \vec{J} & = & \eps A \vec{J}.
\eq
The right-hand side vanishes if $\vec{J}$ is in the kernel of $A$.
This is what happens on the hypersurface $x=1$: Although $A \neq 0$ we have
\bq
 \left. \frac{\partial \vec{J}}{\partial y} \right|_{x=1} & = & 0.
\eq
It follows that the master integrals are constant on the hypersurface $x=1$.
This is a significant simplification for solving the differential equations.


\section{Analytical results}
\label{sect:analytical_results}

The analytic result for the master integrals at a point $(x,y)$ 
is obtained from the value of the master integrals at a boundary point $(x_i,y_i)$
by integrating the system of differential equations along a path from $(x_i,y_i)$ to $(x,y)$.
The result does not depend on the chosen path, but only on the homotopy class of the path.
For the case at hand we have several significant simplification: 
\begin{enumerate}
\item All master integrals are constant on the hypersurface
$x=1$ (corresponding to $p^2=0$ and $m_W^2=m_t^2$)
and we may take the values of the master integrals on this hypersurface as boundary values.
We therefore have a boundary line.
\item It is then sufficient to integrate the differential equation along a straight line from $(1,y)$ to $(x,y)$
with $y=\mathrm{const}$. The polynomials $p_1$-$p_3$ and $p_7$-$p_{16}$ are all linear in $x$ (the polynomials $p_4$-$p_6$
don't contribute along $y=\mathrm{const}$), avoiding the need to factorise higher-order polynomials in $x$.
\item The boundary values on the hypersurface $x=1$ are particularly simple:
$35$ out of the $39$ master integrals vanish on this hypersurface, the four non-vanishing master integrals 
are products of one-loop integrals.
\end{enumerate}
Since all integration kernels are $\mathrm{dlog}$-forms, the result can be expressed in terms of multiple polylogarithms.
Multiple polylogarithms are defined for $l_k \neq 0$ by \cite{Goncharov_no_note,Goncharov:2001,Borwein,Moch:2001zr}
\bq
 \label{Gfuncdef}
 G(l_1,l_2,...,l_k;z)
 & = &
 \int\limits_0^z \frac{dz_1}{z_1-l_1}
 \int\limits_0^{z_1} \frac{dz_2}{z_2-l_2} ...
 \int\limits_0^{z_{k-1}} \frac{dz_k}{z_k-l_k}.
\eq
We have
\bq
\label{Grecursive}
 G(l_1,l_2,...,l_k;z) 
 & = & 
 \int\limits_0^z \frac{dz_1}{z_1-l_1} G(l_2,...,l_k;z_1).
\eq
We define $G(0,...,0;z)$ with $k$ zeros for $l_1$ to $l_k$ to be
\bq
\label{trailingzeros}
 G(0,...,0;z) 
 & = & 
 \frac{1}{k!} \left( \ln z \right)^k.
\eq
This permits us to allow trailing zeros in the sequence
$(l_1,...,l_k)$ by defining the function $G$ with trailing zeros via eq.~(\ref{Grecursive}) 
and eq.~(\ref{trailingzeros}).
For $l_k \neq 0$ we have the scaling relation
\bq
\label{G_scaling_relation}
 G(l_1,...,l_k;z) 
 & = & 
 G(\lambda l_1, ..., \lambda l_k; \lambda z).
\eq
Let us introduce the short-hand notation
\bq
\label{Gshorthand}
 G_{m_1,...,m_k}(l_1,...,l_k;z)
 & = &
 G(\underbrace{0,...,0}_{m_1-1},l_1,...,l_{k-1},\underbrace{0...,0}_{m_k-1},l_k;z),
\eq
where all $l_j$ for $j=1,...,k$ are assumed to be non-zero.
The sum representation of multiple polylogarithms is defined by
\bq 
\label{def_multiple_polylogs_sum}
 \mathrm{Li}_{m_1,...,m_k}(x_1,...,x_k)
  & = & \sum\limits_{n_1>n_2>\ldots>n_k>0}^\infty
     \frac{x_1^{n_1}}{{n_1}^{m_1}}\ldots \frac{x_k^{n_k}}{{n_k}^{m_k}},
\eq
and related to the integral representation by
\bq
\label{Gintrepdef}
 \mathrm{Li}_{m_1,...,m_k}(x_1,...,x_k)
 & = & 
 (-1)^k 
 G_{m_1,...,m_k}\left( \frac{1}{x_1}, \frac{1}{x_1 x_2}, ..., \frac{1}{x_1...x_k};1 \right),
 \nonumber \\
 G_{m_1,...,m_k}(l_1,...,l_k;z) 
 & = & 
 (-1)^k \; \mathrm{Li}_{m_1,...,m_k}\left(\frac{z}{l_1}, \frac{l_1}{l_2}, ..., \frac{l_{k-1}}{l_k}\right).
\eq
The master integrals $J_1$, $J_2$, $J_5$ and $J_6$ are products of one-loop integrals and rather simple.
They are given by
\bq
 J_1 & = &
 e^{2 \gamma_E \eps} \left( \Gamma\left(1+\eps\right) \right)^2,
 \nonumber \\
 J_2 & = &
 e^{2 \gamma_E \eps} \left( \Gamma\left(1+\eps\right) \right)^2 w^{-\eps},
 \nonumber \\
 J_5 & = &
 e^{2 \gamma_E \eps} \frac{\left( \Gamma\left(1+\eps\right) \right)^2 \left( \Gamma\left(1-\eps\right) \right)^2}{\Gamma\left(1-2\eps\right)}  \left(-v\right)^{-\eps},
 \nonumber \\
 J_6 & = &
 e^{2 \gamma_E \eps} \frac{\left( \Gamma\left(1+\eps\right) \right)^2 \left( \Gamma\left(1-\eps\right) \right)^2}{\Gamma\left(1-2\eps\right)}  \left(-v\right)^{-\eps} w^{-\eps}.
\eq
These are the only master integrals, which do not vanish on the hypersurface $x=1$.
All other master integrals vanish on the hypersurface $x=1$.
This fully specifies the boundary conditions on the line $(x,y)=(1,y)$.
Note that the integrals $J_5$ and $J_6$ have logarithmic singularities at $x=1$.
This is most easily seen from eq.~(\ref{variable_trafo}) and the comments made there.
The hypersurface $x=1$ corresponds to the point $(v,w)=(0,1)$ and $J_5$ and $J_6$ have logarithmic singularities at $v=0$.

For the integration in $(x,y)$-space from $(1,y)$ to $(x,y)$ it is better to change variables
from $x$ to $x'=1-x$.
We therefore integrate in $(x',y)$-space from $(0,y)$ to $(x',y)$, where $y$ is treated as a parameter.
This integration
gives multiple polylogarithms of the form
\bq
 G\left(l_1',...,l_k';x'\right),
\eq
where the letters $l_1'$, ..., $l_k'$ are from the alphabet
\bq
 {\mathcal A}
 & = &
 \left\{ 
  0, 1, 2, x_7', x_8', x_9', x_{10}', x_{11}', x_{12}', x_{13}', x_{14}', x_{15}', x_{16}' \right\}.
\eq
The non-trivial letters $x_7'$-$x_{16}'$ are given by
\begin{align}
 x_7'
 & = 
 \frac{2}{1+y},
 &
 x_8'
 & = 
 \frac{1-y}{1+y},
 &
 x_9'
 & = 
 \frac{1+y}{2y},
 \nonumber \\
 x_{10}'
 & = 
 \frac{1+4y-y^2}{\left(1+y\right)^2},
 &
 x_{11}'
 & = 
 \frac{2 \left(1+y^2\right)}{\left(1+y\right)^2},
 &
 x_{12}'
 & = 
 \frac{4y}{\left(1+y\right)^2},
 \nonumber \\
 x_{13}'
 & = 
 - \frac{1-4y-y^2}{2 y \left(1+y\right)},
 &
 x_{14}'
 & = 
 \frac{1+4y-y^2}{2 y \left(1+y\right)},
 &
 x_{15}'
 & = 
 \frac{1-4y-y^2}{\left(1+y\right) \left(1-3y\right)},
 \nonumber \\
 x_{16}'
 & = 
 - \frac{4 y}{\left(1+y\right) \left(1-3y\right)}.
 &&&&
\end{align}
For all basis integrals we write
\bq
 J_k
 & = &
 \sum\limits_{j=0}^\infty \eps^j J_k^{(j)}.
\eq
The quantities $J_k^{(j)}$ are given for $1 \le k \le 39$ and $0 \le j \le 4$
in the supplementary electronic file attached to the arxiv version of this article.
To give an example, let us show the first non-vanishing term of the most complicated integral, the non-planar vertex
correction $J_{39}$.
We find
\bq
\lefteqn{
 J_{39}^{(3)}
 = 
 8\,G\left( 0, 1, 1; x' \right)
 -4\,G\left( 0, 1, x_{8}'; x' \right)
 -4\,G\left( 0, x_{8}', 1; x' \right)
 +4\,G\left( 0, x_{8}', x_{9}'; x' \right)
 +4\,G\left( 0, x_{9}', x_{8}'; x' \right)
 } & &
 \nonumber \\
 & &
 -8\,G\left( 1, 2, 1; x' \right)
 +8\,G\left( 1, x_{15}', 1; x' \right)
 -4\,G\left( 1, x_{15}', x_{8}'; x' \right)
 -4\,G\left( 1, x_{15}', x_{9}'; x' \right)
 \nonumber \\
 & &
 -16\,G\left( 1, x_{7}', 1; x' \right)
 +8\,G\left( 1, x_{7}', x_{8}'; x' \right)
 +12\,G\left( 1, x_{7}', x_{9}'; x' \right)
 +4\,G\left( 1, x_{8}', 1; x' \right)
 \nonumber \\
 & &
 -4\,G\left( 1, x_{8}', x_{9}'; x' \right)
 -4\,G\left( 1, x_{9}', x_{8}'; x' \right)
 +8\,G\left( x_{14}', 2, 1; x' \right)
 -8\,G\left( x_{14}', x_{15}', 1; x' \right)
 \nonumber \\
 & &
 +4\,G\left( x_{14}', x_{15}', x_{8}'; x' \right)
 +4\,G\left( x_{14}', x_{15}', x_{9}'; x' \right)
 +16\,G\left( x_{14}', x_{7}', 1; x' \right)
 -8\,G\left( x_{14}', x_{7}', x_{8}'; x' \right)
 \nonumber \\
 & &
 -12\,G\left( x_{14}', x_{7}', x_{9}'; x' \right)
 -4\,G\left( x_{14}', x_{8}', 1; x' \right)
 +4\,G\left( x_{14}', x_{8}', x_{9}'; x' \right)
 +4\,G\left( x_{14}', x_{9}', x_{8}'; x' \right)
 \nonumber \\
 & &
 +8\,G\left( x_{15}', 1, 1; x' \right)
 +4\,G\left( x_{15}', 1, x_{8}'; x' \right)
 -8\,G\left( x_{15}', 2, 1; x' \right)
 +4\,G\left( x_{15}', x_{14}', x_{8}'; x' \right)
 \nonumber \\
 & &
 +4\,G\left( x_{15}', x_{14}', x_{9}'; x' \right)
 +4\,G\left( x_{15}', x_{7}', 1; x' \right)
 -12\,G\left( x_{15}', x_{7}', x_{8}'; x' \right)
 -8\,G\left( x_{15}', x_{7}', x_{9}'; x' \right)
\hspace*{9mm}
 \nonumber \\
 & &
 -4\,G\left( x_{15}', x_{8}', 1; x' \right)
 +4\,G\left( x_{15}', x_{8}', x_{9}'; x' \right)
 +4\,G\left( x_{15}', x_{9}', x_{8}'; x' \right)
 -8\,G\left( x_{7}', 1, 1; x' \right)
 \nonumber \\
 & &
 +4\,G\left( x_{7}', 1, x_{8}'; x' \right)
 +4\,G\left( x_{7}', x_{8}', 1; x' \right)
 -4\,G\left( x_{7}', x_{8}', x_{9}'; x' \right)
 -4\,G\left( x_{7}', x_{9}', x_{8}'; x' \right).
\eq


\section{Numerical results}
\label{sect:numerical_results}

Of particular interest are numerical results for 
\bq
 p^2 & = & m_H^2.
\eq
Since $p^2>0$, we are not in the Euclidean region. 
Feynman's $i0$-prescription instructs us to take a small imaginary part into account: $p^2 \rightarrow p^2 + i0$.
This selects the correct branches for the two square roots $\sqrt{-v(4-v)}$ and $\sqrt{\lambda(v,w,1)}$.
With
\bq
 m_W \; = \; 80.38 \; \mathrm{GeV},
 \;\;\;\;\;\;
 m_H \; = \; 125.2 \; \mathrm{GeV},
 \;\;\;\;\;\;
 m_t \; = \; 173.1 \; \mathrm{GeV}
\eq
we obtain for the variables $x$ and $y$
\bq
 x 
 \; = \;
 0.7384 + 0.6743 i,
 & &
 y
 \; = \;
 0.3987 i.
\eq
\begin{table}[!htbp]
\begin{center}
{\scriptsize
\begin{tabular}{|l|lllll|}
 \hline 
 & $\eps^0$ & $\eps^1$ & $\eps^2$ & $\eps^3$ & $\eps^4$ \\
 \hline 
$J_{ 1}$ & $        1$ & $        0$ & $ 1.6449341$ & $-0.80137127$ & $ 1.8940657$ \\ 
$J_{ 2}$ & $        1$ & $ 1.5342081$ & $ 2.8218314$ & $ 2.3241685$ & $ 2.8313617$ \\ 
$J_{ 3}$ & $        0$ & $-0.74005414 i$ & $-0.069477641 i$ & $-1.2212434 i$ & $ 0.47861551 i$ \\ 
$J_{ 4}$ & $        0$ & $-0.74005414 i$ & $-1.2048747 i$ & $-2.1988043 i$ & $-1.9222094 i$ \\ 
$J_{ 5}$ & $        1$ & $ 0.64791401 +  3.1415927 i$ & $-4.7249059 +  2.0354819 i$ & $-6.357481 -4.5083042 i$ & $-2.2935901 -13.276148 i$ \\ 
$J_{ 6}$ & $        1$ & $ 2.1821222 +  3.1415927 i$ & $-2.5539737 +  6.8553389 i$ & $-12.242073 +  2.3118807 i$ & $-16.987211 -15.906447 i$ \\ 
$J_{ 7}$ & $        0$ & $ 0.74005414 i$ & $-1.2490335 i$ & $ 4.0114542 i$ & $-7.931414 i$ \\ 
$J_{ 8}$ & $        0$ & $        0$ & $-0.54768013$ & $ 0.47549335$ & $-2.5261306$ \\ 
$J_{ 9}$ & $        0$ & $ 1.4585842 i$ & $ 0.050868134 i$ & $ 6.4537259 i$ & $-4.1565512 i$ \\ 
$J_{ 10}$ & $        0$ & $ 1.5342081$ & $-1.7502763$ & $ 5.7981619$ & $-12.790768$ \\ 
$J_{ 11}$ & $        0$ & $        0$ & $ 2.2406312$ & $ 1.5008239$ & $ 10.399293$ \\ 
$J_{ 12}$ & $        0$ & $-1.5342081$ & $ 0.91070302$ & $-5.9968338$ & $ 8.8611857$ \\ 
$J_{ 13}$ & $        0$ & $        0$ & $-0.54768013$ & $-0.10283443$ & $-0.91150188$ \\ 
$J_{ 14}$ & $        0$ & $ 0.74005414 i$ & $-2.3249487 +  0.54896908 i$ & $-1.7246372 -3.4477675 i$ & $ 3.1827067 -5.0304701 i$ \\ 
$J_{ 15}$ & $        0$ & $        0$ & $-0.40473314$ & $-0.19458947$ & $-0.73218618$ \\ 
$J_{ 16}$ & $        0$ & $-1.4801083 i$ & $ 2.8153204 i$ & $-7.8872093 i$ & $ 15.843952 i$ \\ 
$J_{ 17}$ & $        0$ & $        0$ & $-0.27384007$ & $ 0.25481947$ & $-1.2423838$ \\ 
$J_{ 18}$ & $        0$ & $        0$ & $        0$ & $ 0.54349879$ & $-0.505106$ \\ 
$J_{ 19}$ & $        0$ & $-0.74005414 i$ & $ 1.3879888 i$ & $-3.8717735 i$ & $ 8.3164903 i$ \\ 
$J_{ 20}$ & $        0$ & $        0$ & $-0.40473314$ & $-0.012836717$ & $-1.766483$ \\ 
$J_{ 21}$ & $        0$ & $        0$ & $        0$ & $ 1.0679293$ & $ 0.50513135$ \\ 
$J_{ 22}$ & $        0$ & $        0$ & $ 0.40473314 +  2.8771124 i$ & $-2.1230218 -1.197403 i$ & $ 0.75622025 +  11.086074 i$ \\ 
$J_{ 23}$ & $        0$ & $        0$ & $        0$ & $ 1.4266372$ & $-0.44333884$ \\ 
$J_{ 24}$ & $        0$ & $        0$ & $-0.80946628$ & $ 0.078302369$ & $-3.4351805$ \\ 
$J_{ 25}$ & $        0$ & $        0$ & $        0$ & $ 1.447072$ & $-0.3662879$ \\ 
$J_{ 26}$ & $        0$ & $        0$ & $-0.83957331$ & $-0.038098053$ & $-3.7104375$ \\ 
$J_{ 27}$ & $        0$ & $        0$ & $ 0.40473314$ & $ 0.19458947 -0.29952444 i$ & $ 0.73218618 -0.17212665 i$ \\ 
$J_{ 28}$ & $        0$ & $        0$ & $        0$ & $ 2.4399925 +  2.5430133 i$ & $-6.9287969 +  2.8702957 i$ \\ 
$J_{ 29}$ & $        0$ & $        0$ & $        0$ & $ 0.69594214$ & $ 1.4888349$ \\ 
$J_{ 30}$ & $        0$ & $        0$ & $        0$ & $ 1.3639124$ & $ 2.6684304$ \\ 
$J_{ 31}$ & $        0$ & $        0$ & $        0$ & $ 1.530613 -0.59904887 i$ & $ 2.5899951 -1.0224563 i$ \\ 
$J_{ 32}$ & $        0$ & $        0$ & $        0$ & $-1.6890882$ & $-0.27250855$ \\ 
$J_{ 33}$ & $        0$ & $        0$ & $        0$ & $ 0.372767$ & $ 0.60475274$ \\ 
$J_{ 34}$ & $        0$ & $        0$ & $        0$ & $ 0.30226015$ & $ 0.61200492$ \\ 
$J_{ 35}$ & $        0$ & $        0$ & $-0.29189318$ & $-0.65744728$ & $-1.3431597$ \\ 
$J_{ 36}$ & $        0$ & $        0$ & $        0$ & $ 1.5567823 -0.61339819 i$ & $ 2.7076661 -1.0872344 i$ \\ 
$J_{ 37}$ & $        0$ & $        0$ & $        0$ & $-0.78991058$ & $ 0.12604664$ \\ 
$J_{ 38}$ & $        0$ & $        0$ & $        0$ & $ 0.050162573$ & $ 0.077279399$ \\ 
$J_{ 39}$ & $        0$ & $        0$ & $        0$ & $ 0.18876826$ & $ 0.41154739$ \\ 
 \hline 
\end{tabular}
}
\end{center}
\caption{
Numerical results for the first five terms of the $\eps$-expansion of the master integrals $J_{1}$-$J_{39}$ 
at the kinematic point $s=m_H^2$.
}
\label{table_numerical_results}
\end{table}
The values of the master integrals at this point are given to $8$ digits in table~\ref{table_numerical_results}.
They are easily computed to arbitrary precision by evaluating the multiple polylogarithms with the help of \verb|GiNaC| \cite{Bauer:2000cp,Vollinga:2004sn}.
In addition we verified the first few digits at various kinematic points
with the help of the programs \verb|sector_decomposition| \cite{Bogner:2007cr}
and
\verb|pySecDec| \cite{Borowka:2017idc,2018arXiv181111720B}.


\section{Conclusions}
\label{sect:conclusions}

In this paper we presented the two-loop master integrals 
relevant to the ${\mathcal O}(\alpha \alpha_s)$-corrections 
to the decay $H \rightarrow b \bar{b}$ through a $H t \bar{t}$-coupling.
We kept the exact dependence of the masses of the heavy particles ($m_W$ and $m_t$) and the momentum $p^2$ of the Higgs boson,
but neglected the $b$-quark mass.
All master integrals are expressed in terms of multiple polylogarithms with an alphabet of $13$ letters.
They can be evaluated to arbitrary precision with the help of the \verb|GiNaC|-library.
For the special case $p^2=m_H^2$ we presented the numerical values.

\subsection*{Acknowledgements}

This work has been supported by the 
Cluster of Excellence ``Precision Physics, Fundamental Interactions, and Structure of Matter'' 
(PRISMA+ EXC 2118/1) funded by the German Research Foundation (DFG) 
within the German Excellence Strategy (Project ID 39083149).

S.W. would like to thank the Institute for Theoretical Studies in Zurich
for hospitality, where part of this work was carried out.
The Feynman diagrams in this article have been made with the program {\tt axodraw} \cite{Vermaseren:1994je}.


\begin{appendix}

\section{Master topologies}
\label{sect:master_topologies}

\begin{figure}[!hbp]
\begin{center}
\includegraphics[scale=1.0]{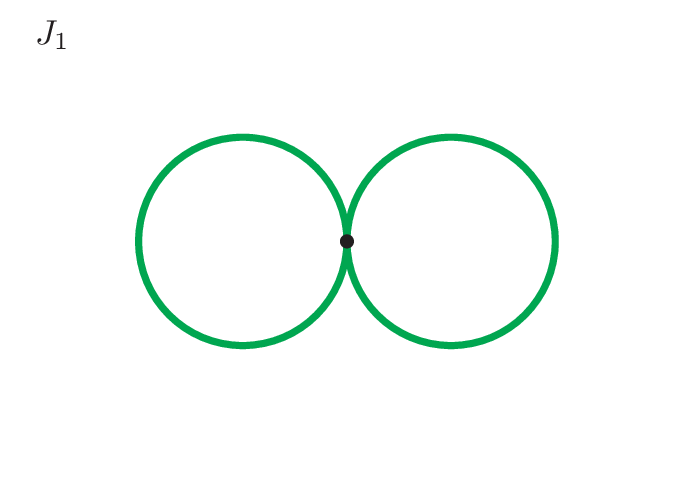}
\includegraphics[scale=1.0]{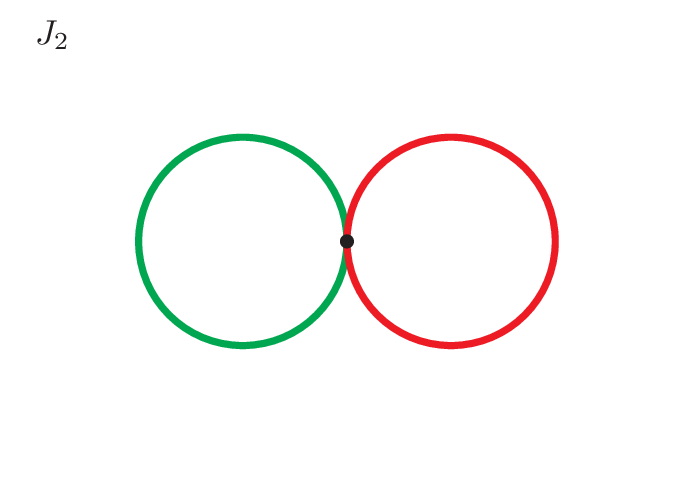}
\end{center}
\caption{
Master topologies (part 1).
}
\label{fig_master_topologies_part1}
\end{figure}

\begin{figure}
\begin{center}
\includegraphics[scale=1.0]{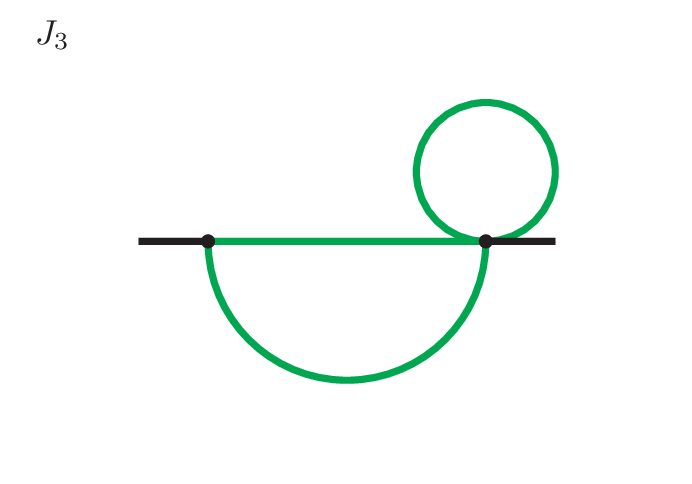}
\includegraphics[scale=1.0]{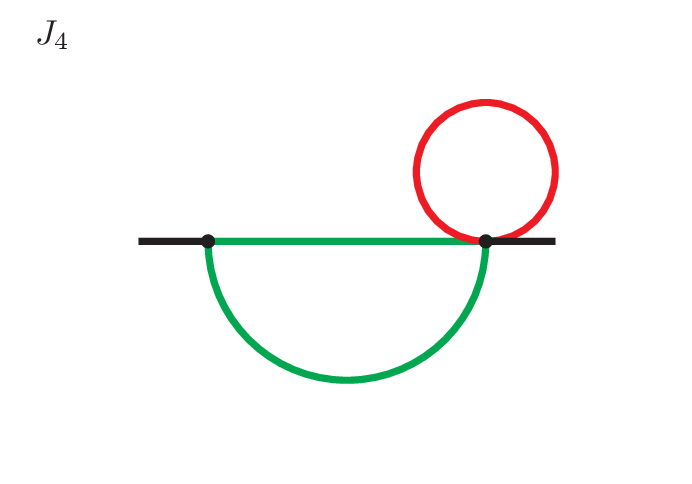}
\includegraphics[scale=1.0]{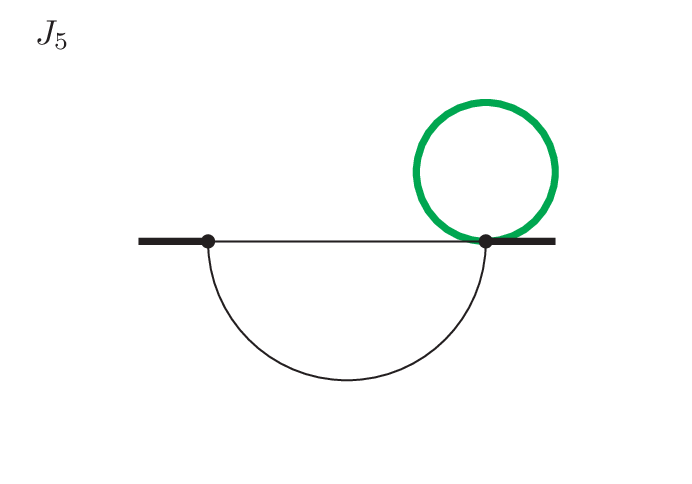}
\includegraphics[scale=1.0]{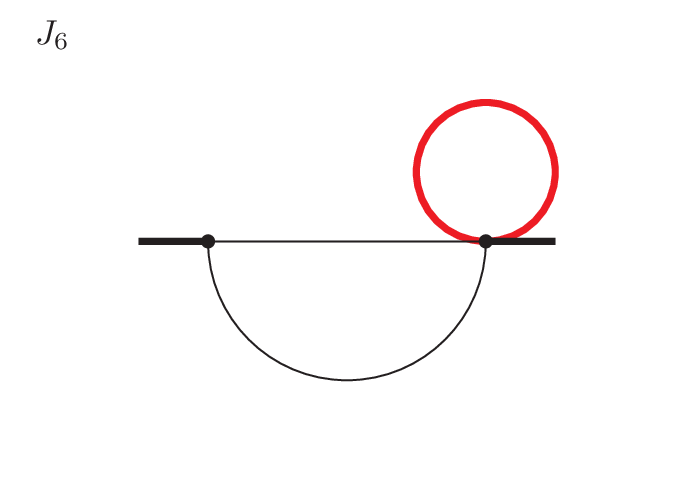}
\includegraphics[scale=1.0]{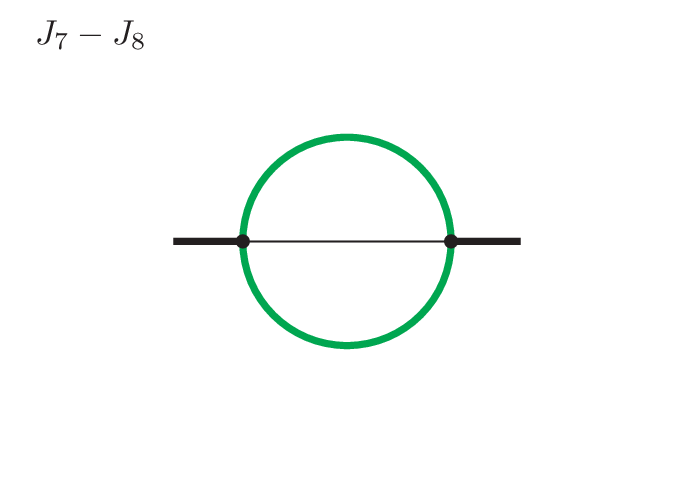}
\includegraphics[scale=1.0]{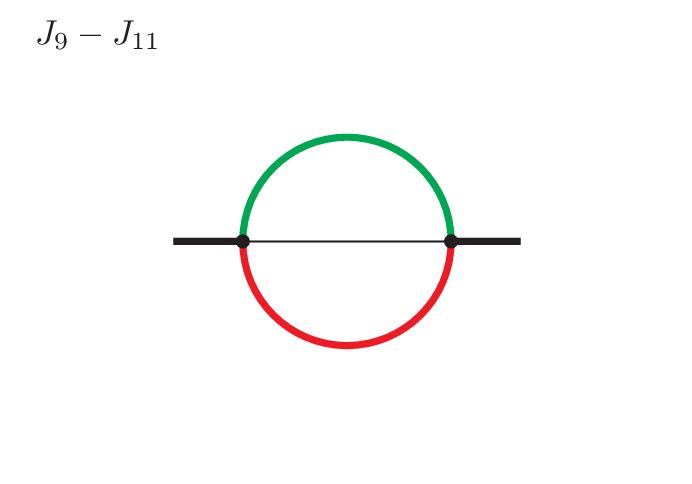}
\includegraphics[scale=1.0]{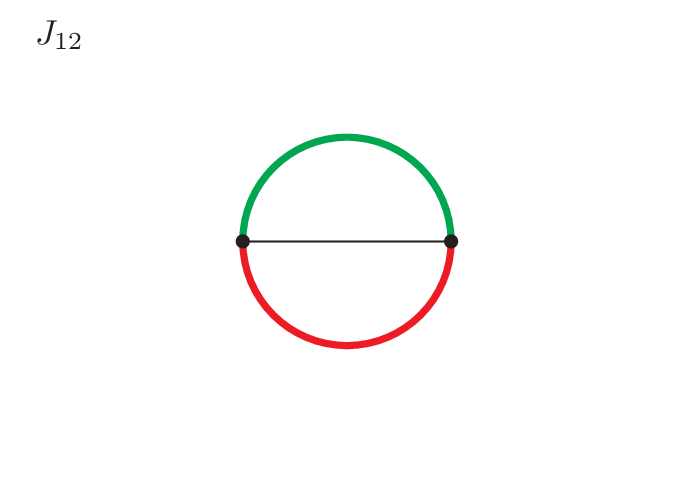}
\includegraphics[scale=1.0]{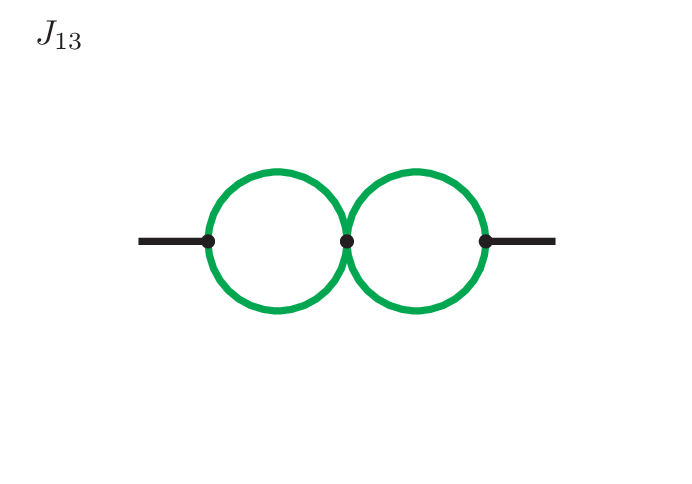}
\end{center}
\caption{
Master topologies (part 2).
}
\label{fig_master_topologies_part2}
\end{figure}

\begin{figure}
\begin{center}
\includegraphics[scale=1.0]{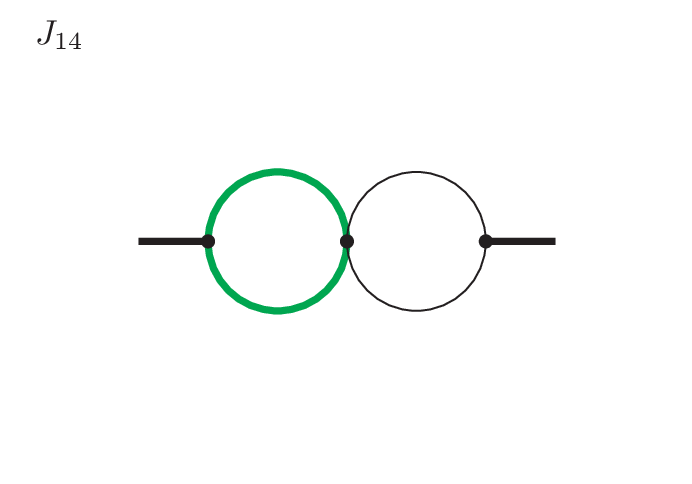}
\includegraphics[scale=1.0]{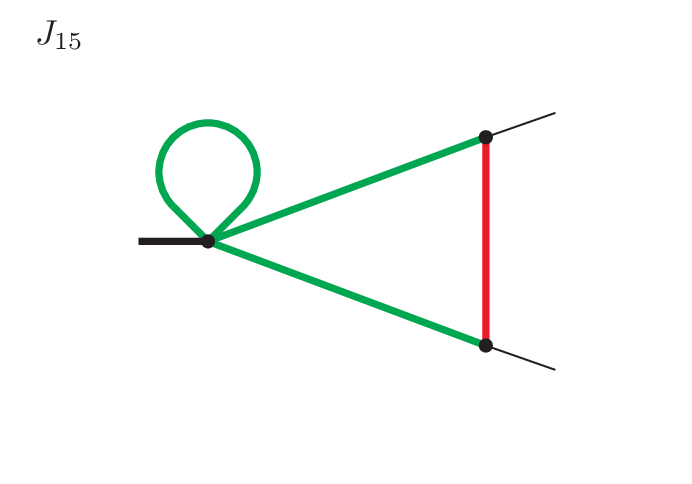}
\includegraphics[scale=1.0]{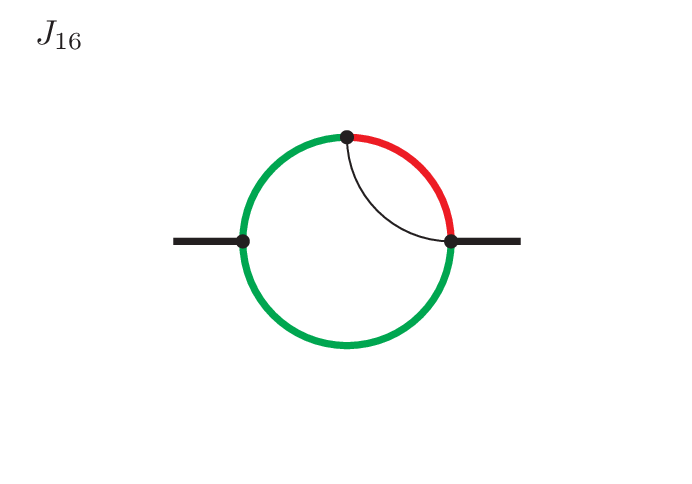}
\includegraphics[scale=1.0]{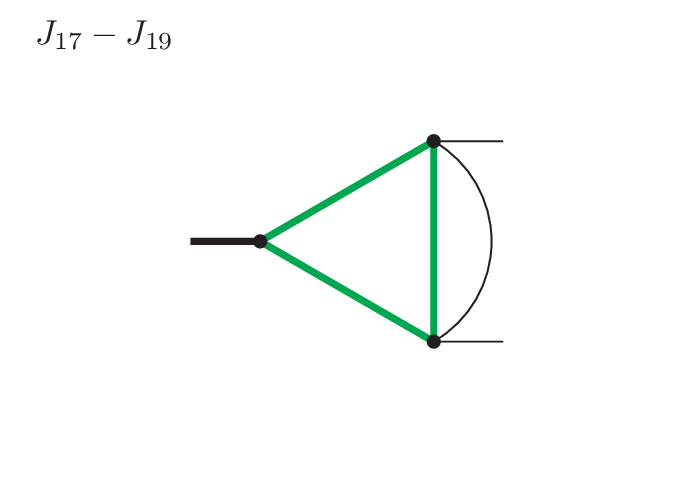}
\includegraphics[scale=1.0]{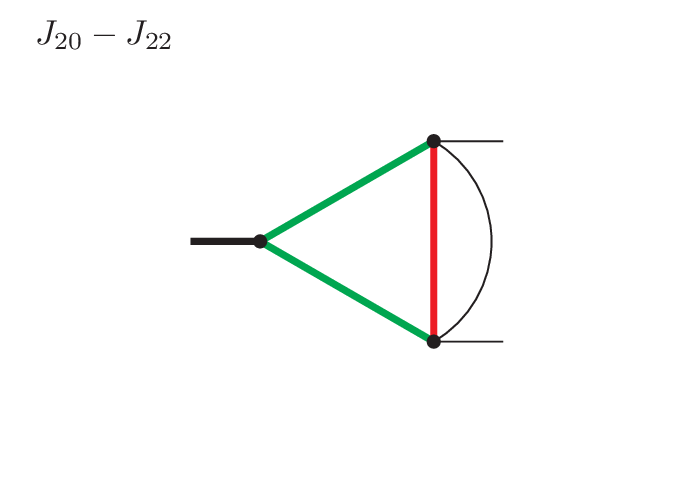}
\includegraphics[scale=1.0]{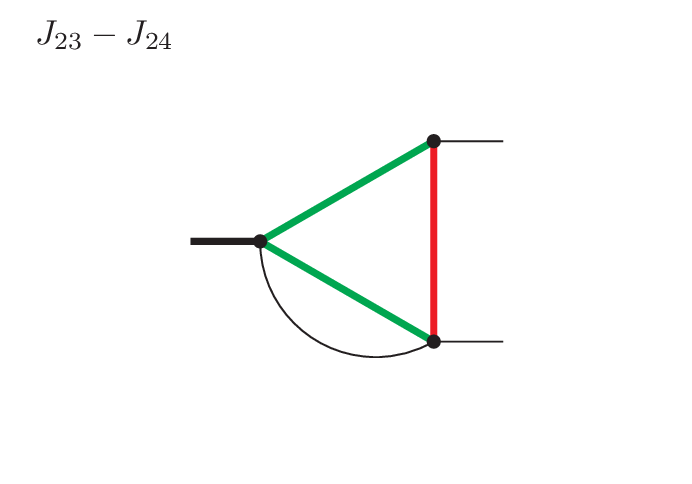}
\includegraphics[scale=1.0]{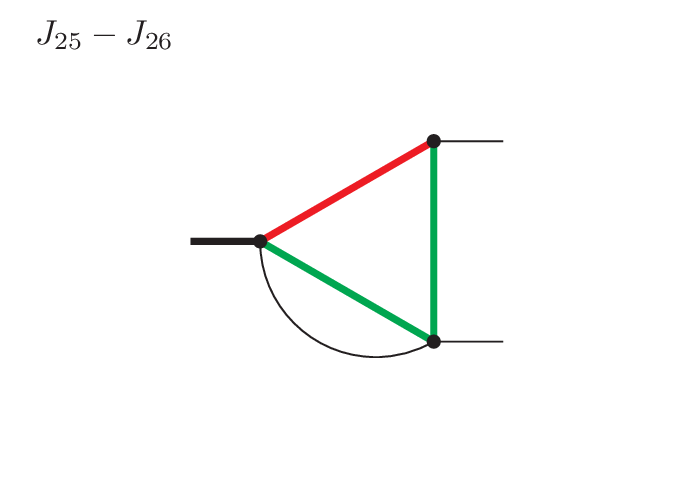}
\includegraphics[scale=1.0]{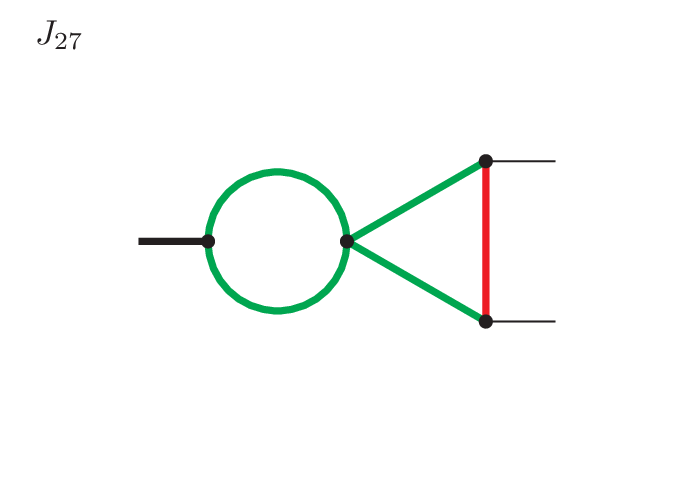}
\end{center}
\caption{
Master topologies (part 3).
}
\label{fig_master_topologies_part3}
\end{figure}

\begin{figure}
\begin{center}
\includegraphics[scale=1.0]{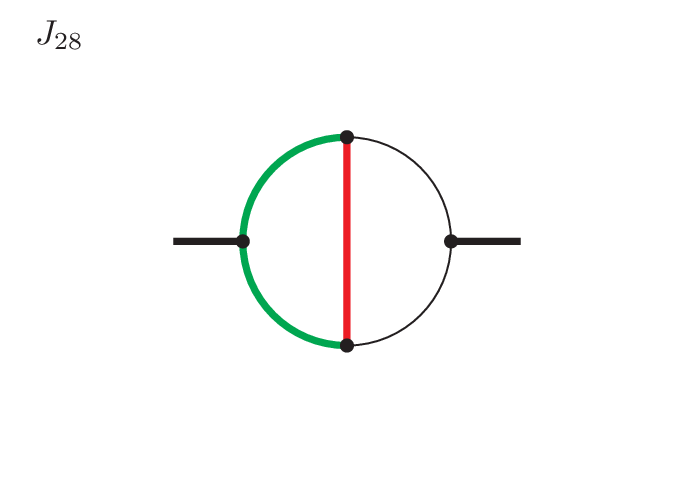}
\includegraphics[scale=1.0]{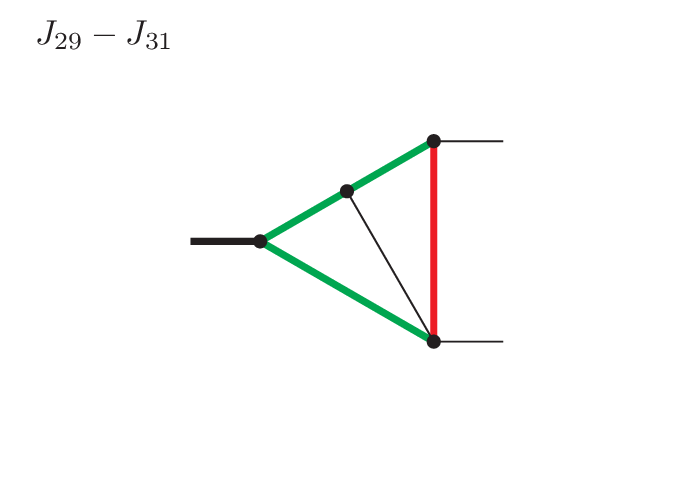}
\includegraphics[scale=1.0]{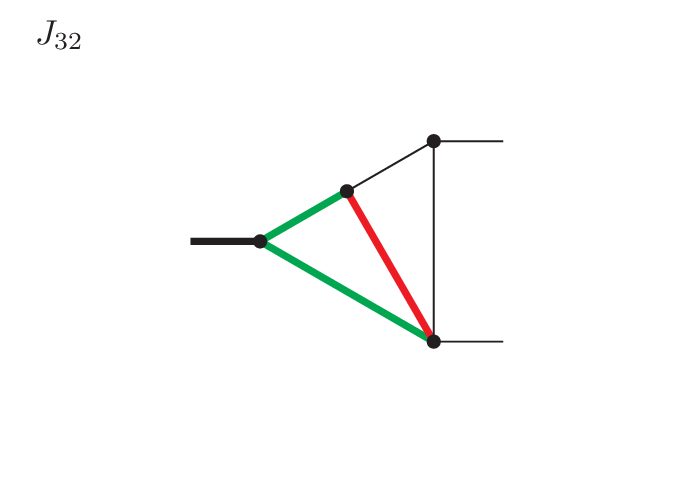}
\includegraphics[scale=1.0]{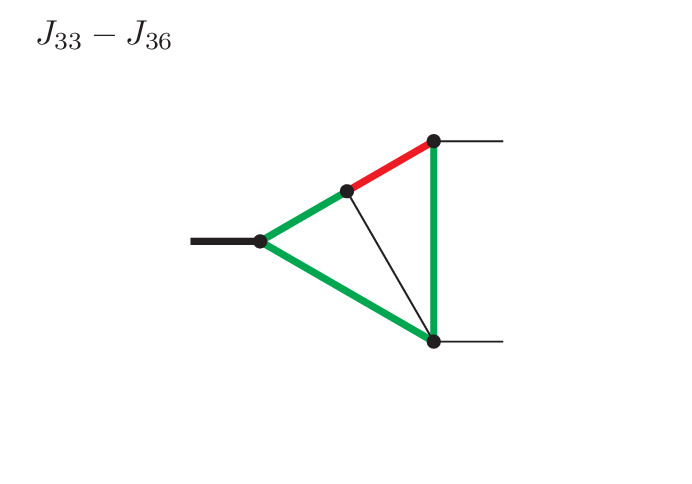}
\includegraphics[scale=1.0]{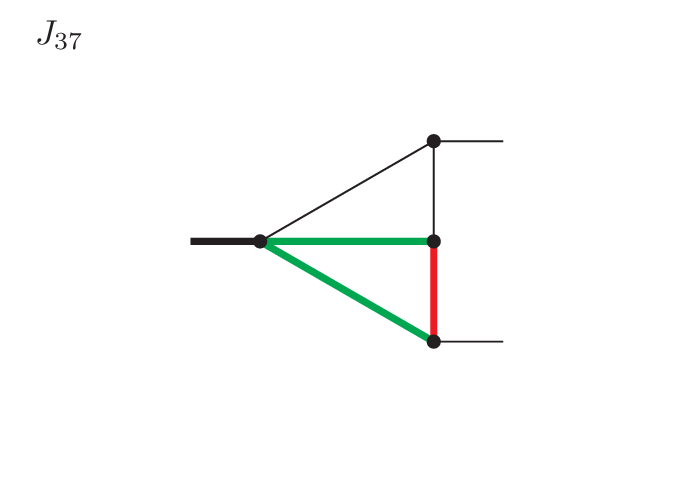}
\includegraphics[scale=1.0]{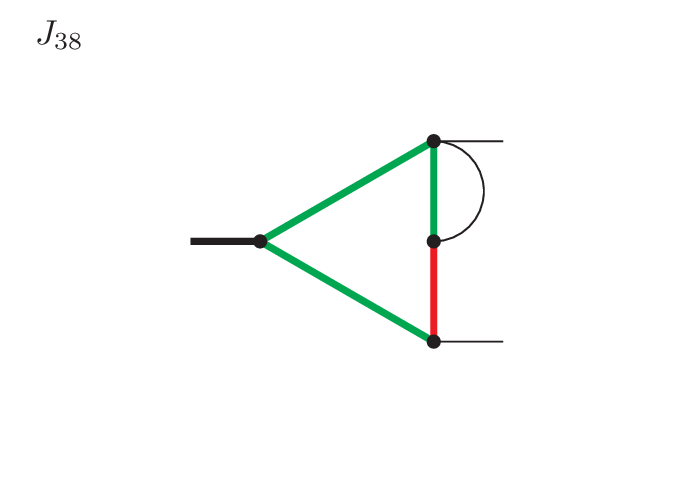}
\includegraphics[scale=1.0]{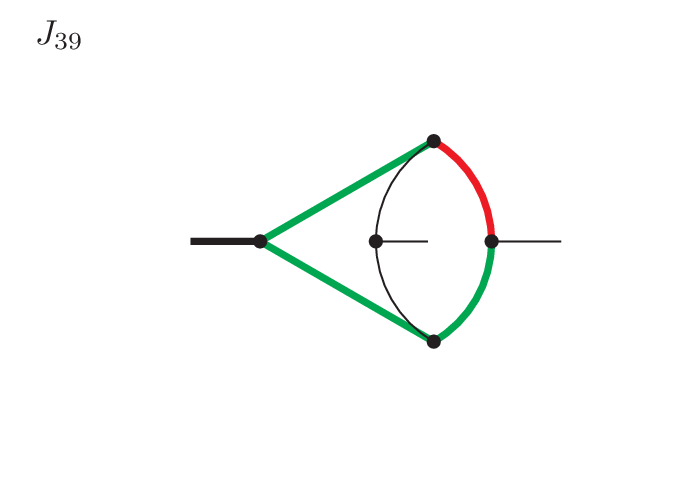}
\end{center}
\caption{
Master topologies (part 4).
}
\label{fig_master_topologies_part4}
\end{figure}

In this appendix we show in tables~\ref{fig_master_topologies_part1}-\ref{fig_master_topologies_part4} the diagrams of all master topologies.
\FloatBarrier


\section{Supplementary material}
\label{sect:supplement}

Attached to the arxiv version of this article is an electronic file in ASCII format with {\tt Maple} syntax, defining the quantities
\begin{center}
 \verb|A|, \; \verb|J|.
\end{center}
The matrix \verb|A| appears in the differential equation
\bq
 d \vec{J} & = & \eps A \vec{J}.
\eq
The entries of the matrix $A$ are ${\mathbb Z}$-linear combinations of $\omega_1$, ..., $\omega_{15}$, defined in eq.~(\ref{diff_forms}).
These differential forms are denoted by
\begin{center}
 \verb|omega_1|, ..., \verb|omega_15|.
\end{center}
The vector \verb|J| contains the results for the master integrals up to order $\eps^4$ in terms of multiple polylogarithms.
The variable $\eps$ is denoted by \verb|eps|, $\zeta_2$, $\zeta_3$, $\zeta_4$ by
\begin{center}
 \verb|zeta_2|, \verb|zeta_3|, \verb|zeta_4|,
\end{center}
respectively. For the notation of multiple polylogarithms we given an example:
$G(x_7',x_8',1;x')$ is denoted by
\begin{center}
 \verb|Glog([xp7,xp8,1],xp)|.
\end{center}

\end{appendix}

{\footnotesize
\bibliography{/home/stefanw/notes/biblio}
\bibliographystyle{/home/stefanw/latex-style/h-physrev5}
}

\end{document}